# Exploiting product molecule number to consider reaction rate fluctuation in elementary reactions

Seong Jun Park[1]

[1] National CRI-Center for Chemical Dynamics in Living Cells, Chung-Ang University, Seoul 06974, Korea.

**Abstract**

In many chemical reactions, reaction rate fluctuation is inevitable. Reaction rates are different whenever chemical reaction occurs due to their dependence on the number of reaction events or the product number. As such, understanding the impact of rate fluctuation on product number counting statistics is of the utmost importance when developing a quantitative explanation of chemical reactions. In this work, we present a master equation that describes reaction rates as a function of product number and time. Our equal reveals the relationship between the reaction rate and product number fluctuation. Product number counting statistics uncovers a stochastic property of the product number; product number directly manipulates the reaction rate. Specifically, we find that product number shows super-Poisson characteristics when the product number increases, inducing an increase in the reaction rate. While, on the other hand, when the product number shows sub-Poisson characteristics with an increase in the product number, this is induced by a decrease in the reaction rate. Furthermore, our analysis exploits reaction rate fluctuation, enabling the quantification of the deviation of an elementary reaction process from a renewal process.

Most chemical reactions involve reaction rate fluctuation which is adaptation and response to changing environments. Several studies have focused on quantifying the rate coefficient in enzyme reactions [1-4] and have shown that this rate coefficient is a random variable, not a constant. Motivated by these observations, Lim.*et.al* [5] introduce the concept of a *vibrant* reaction process with the rate coefficient modeled as a stochastic variable coupled to reaction environments, proposing a method for considering rate coefficient fluctuations. Extending the concept to single enzyme reactions, Park. *et.al* [6] suggest a new statistical kinetics that characterizes the dynamics of enzyme activity. Recently, the Chemical Fluctuation Theorem (CFT) governing general birth-death process was recently shown to successfully and quantitatively provide an accurate description of gene expression [7]. These examples make it is obvious that rate coefficient fluctuates due to reaction environment and differs whenever reaction occurs, demonstrating that reaction rate varied with the number of reaction events, or product number.

The master equation[8] is perhaps the most common approach used to explain product number fluctuation caused by reaction events. The conventional master equation assumes that the rate coefficient of elementary reactions are constants that remain unchanged, regardless of product number, and under a homogeneous reaction environment, the conventional master equation is, in fact, applicable because the rate coefficient does indeed remain constant. However, in a heterogeneous reaction environment, this rate coefficient does not remain constant because it varies with the product number, or the number of reaction events in an elementary reaction.

In the current work, we propose an alternative description for reaction rate fluctuation induced by reaction events or product number and present our new theoretical results. By describing reaction rates as a function of product number and time, we obtain a new master equation, and we derive the mean and variance of the product number from this new equation. Our result yields the relationship between the fluctuating reaction rate and product number counting statistics. We are additionally able to obtain the probability density function of the time required to produce *n* product molecules. Using this probability density function, we are also able to estimate the deviation of an elementary reaction from a renewal process as the product number increases.

To consider a realistic reaction system that cannot be accurately described by the conventional master equation, we first derive a new type of master equation for an elementary reaction, $A \to B$. Our equation defines a reaction rate coefficient that fluctuates with time and product number; that is, in this definition, our reaction rate coefficient is a function of time and product number. We begin with two assumptions. We first assume that within the sufficiently small time interval, $h(h>0)$, the probability of a single reaction occurring can be approximated as $\lambda_n(t)h$ when the product number is *n* at time *t*, where $\lambda_n(t)$ is a real non-negative function and represents the reaction rate. Our second assumption is that the probability of two or more reactions occurring in time interval $h$ is zero. We can then denote the probability of the product number being *n* at time *t* by $P_n(t)$, which satisfies the normalization condition, $\sum_{n=0}^{\infty} P_n(t) = 1$. With these assumptions in place, we can approximate the probability of the product number being *n* at time *t+h* by multiplying the probability of the product number being *n* at time *t* by the probability of no reaction occurring in the time interval, (*t, t+h*), and then multiply the probability of the product number being *n*-1 at time *t* by the probability of a single reaction occurring in time interval (*t, t+h*). This results in

$$P_n(t+h) \approx P_n(t)\left(1 - \lambda_n(t)h\right) + P_{n-1}(t)\lambda_{n-1}(t)h, \tag{1}$$

owing to the fact that the probability of a single reaction occurring in (*t, t+h*) is approximately equal to $\lambda_n(t)h$ when the product number is equal to *n* at time *t*. Taking limit as $h \to 0$, the resulting differential equation is

$$\frac{\partial P_n(t)}{\partial t} = P_{n-1}(t)\lambda_{n-1}(t) - P_n(t)\lambda_n(t). \tag{2}$$

This formulation of our master equation clarifies its physical meaning. This equation is a gain-loss equation about the product number. The first term on the right-hand side (R.H.S) of Eq. (2) is a gain-term representing a reaction event when the product number increases from *n*-1, while the second term is a loss-term that represents a reaction event when the product number is *n*. Eq. (2) reduces to the conventional master equation when $\lambda_n(t)$ is constant.

Eq. (2) enables us to compute $P_n(t)$ even when only the reaction rate and its initial condition, $P_n(0)$, are known. It is worth mentioning that, at this stage, $P_n(t)$ can be represented only by a function of the reaction rate, $\lambda_n(t)$, which is remarkable because the relationship between $P_n(t)$ and $\lambda_n(t)$ was not well understood until now, excluding the case when $\lambda_n(t)$ is constant or only function of time. For more details on this matter, see Supplementary Method I.

Using Eq. (2) to examine the counting statistics of the product number, we can take the first moment of the product number, given by $\langle n(t) \rangle = \sum_{n=0}^{\infty} nP_n(t)$. We can then obtain

$$\langle n(t) \rangle = \int_0^t \langle \lambda_n(\tau) \rangle d\tau, \tag{3}$$

where $\langle \lambda_n(t) \rangle = \sum_{n=0}^{\infty} \lambda_n(t) P_n(t)$. Taking the derivative of Eq. (3) with respect to *t*, we then obtain a result in good agreement with our expectation, which is that a change in the mean product number with respect to time is the same as the mean of the reaction rate, $\langle \lambda_n(t) \rangle$.

Product number fluctuation can be exploited to obtain valuable information about the stochastic property of a reaction process. We can extract the stochastic property of an elementary reaction process by analyzing the variance of the product number, $\sigma^2(t) = \langle n^2(t) \rangle - \langle n(t) \rangle^2$. Similarly, one can discover the second moment, $\langle n^2(t) \rangle$, of the product number and obtain the Mandel's Q parameter of the product number as follows,

$$Q_n(t) = \frac{2}{\langle n(t) \rangle} \int_0^t \rho(\tau) \sigma_n(\tau) \sigma_{\lambda_n}(\tau) \, d\tau \tag{4}$$

where $Q_n(t) = \sigma_n^2(t)/\langle n(t) \rangle - 1$ and $\rho(t)$ $(|\rho(t)| \leq 1)$ are the correlation coefficient of the product number, $n$, and the reaction rate, $\lambda_n(t)$, respectively. $\sigma_n(t)$ and $\sigma_{\lambda_n}(t)$ are the standard deviations of the product number and the reaction rate, respectively. For detailed derivations of Eqs. (3) and (4), see Supplementary Method II. From Eq. (4), we can note that the correlation coefficient, $\rho(t)$, of the product number, $n$, and the reaction rate, $\lambda_n(t)$, determine the product number's stochastic property. When the reaction rate is not correlated with the product number or when the reaction rate shows no dependence on the product number, the product number counting statistics show Poisson characteristics, $Q_n(t) = 0$, due to the fact that $\rho(t) = 0$. We confirm this to be the case in Supplementary Method III. When the reaction rate, $\lambda_n(t)$, is positively correlated with the product number or when $\lambda_n(t)$ is an increasing function of the product number, the counting statistics of the product number bear a super-Poisson character, $Q_n(t) > 0$, (Fig. 1(a)), while on the other hand, when $\lambda_n(t)$ is inversely correlated with the product number or when $\lambda_n(t)$ is a decreasing function of the product number, we instead see sub-Poisson characteristics, $Q_n(t) < 0$, due to the fact that $\rho(t) < 0$

(Fig. 1(b)). It is important because when the reaction rate is an oscillating function of the product number, the product number counting statistics can be Poisson, super-Poisson, or sub-Poisson. For this case, it is necessary to calculate the definite integral on the right-hand side of Eq. (4) in order for us to determine which product number fluctuation shows stochastic properties (Fig. 1(c)). Further details on a faster mean by which to obtain Eqs. (3) and (4) can be found in Supplementary Method IV. Comparing with Eq. (4) and the chemical fluctuation theorem gives the relation between the time correlation function (TCF) of the production rate, and the probability density function of reaction time interval is obtained (Supplementary Method V).

It is well-known that, at long times, the Mandel's Q parameter for the product number produced by a renewal reaction process is the same as the randomness parameter (the difference between the variance over the mean square and unity) in the waiting time required to produce one molecule [9]. When we determine how an elementary reaction process deviates from a renewal process, the important quantity we must pay special attention to is the probability density function of the time required to produce $n$ product molecules, which we denote as $\varphi_n(t)$ for $n = 1, 2, 3, \cdots$. The non-renewal property is defined mathematically as $\hat{g}_n(s) = \hat{\varphi}_n(s) / \hat{\varphi}_1(s)^n$ for $n=1,2,3,\cdots$. $\hat{f}(s)$ denotes the Laplace transform of $f(t)$. $\hat{g}_n(s) = 1$ for a renewal process. In other words, $\hat{g}_n(s) \neq 1$ signifies a non-renewal process.

We are now also able to estimate the deviation of an elementary reaction from a renewal process as the product number increases. To put it another way, we can express $\hat{g}_n(s)$ as a function of the reaction rate, $\lambda_n(t)$. We need the relationship between $P_n(t)$ and $\varphi_n(t)$ in order to establish that $\hat{g}_n(s)$ is a function of the reaction rate, $\lambda_n(t)$. For an in-depth explanation of the relationship between $P_n(t)$ and $\varphi_n(t)$, see Supplementary Method IV. We

can then use Eq. (M4-12) from Supplementary Information to rewrite $\hat{g}_n(s)$ as

$$\hat{g}_n(s) = \frac{1 - s\hat{z}_{n-1}(s)}{\left(1 - s\hat{z}_0(s)\right)^n}, \tag{5}$$

where $z_n(t) = e^{-r_n(t)}\left(1 + q_n(t)\right)$, $r_n(t) = \int_0^t \lambda_n(\tau)d\tau$, $q_n(t) = \int_0^t \left(1 + q_{n-1}(\tau)\right)\lambda_n(\tau)e^{r_n(\tau) - r_{n-1}(\tau)}d\tau$, and $q_0(t) = 0$ for $n=1,2,3,\cdots$. $z_n(t)$ can be interpreted as the probability of the product number being less than or equal to $n$ in the time interval [0, $t$] (see Supplementary Method IV). Let us refer to $\hat{g}_n(s) - 1$ as a "non-renewal quotient" that approaches zero when an elementary reaction gets closer to a renewal process. We can then say that a chemical reaction is a non-renewal process when the non-renewal quotient is not equal to zero. We confirm this by investigating three models of $\lambda_n(t)$ in Fig 2, and as can be easily seen, as the product number increases, the non-renewal quotient becomes farther and farther from zero. In addition, the nonrenewal quotient deviates farther from zero as the Laplace variable, $s$, in $\hat{g}_n(s)$ becomes larger with the fixed number of product molecules, $n$. The large Laplace variable $s$ corresponds to short times in the time domain, while the small Laplace variable $s$ corresponds to long times in the time domain. For the three given models in Fig. 2, the non-renewal character becomes more pronounced when the waiting time required to produce product molecules is shorter. We are then enabled to quantify the deviation of an elementary reaction from a renewal process even if only the reaction rate, $\lambda_n(t)$, is known.

We leave for future research studying a birth-death process with reaction rates fluctuating with product number. Birth-death processes are relevant to many fields that study the populations of systems, such as queueing models and gene expression [10-12]. The general results derived in this work serve as useful tools for investigating birth-death processes.

In summary, the current work suggests a method which considers fluctuation in the reaction rate by making use of the product number. In order to do so, we develop new type of master equation for elementary reactions, and using this equation, we investigate the product number counting statistics and the nonrenewal quotient of an elementary reaction. We find that when the reaction rate increases with the product number, the product number counting statistics shows super-Poissonian characteristics, while a decrease in the product number results in the product number counting statistics showing sub-Poissonian characteristics. If the reaction rate is known, we can use the non-renewal quotient developed here and determine the extent to which an elementary reaction can be defined as a non-renewal process. The ability to determine the stochastic property of the product number fluctuation, Eq. (4), and nonrenewal quotient, Eq. (5), when only the reaction rate is known is a significant achievement that expands our understanding of elementary reactions.

## Acknowledgement

The authors are grateful to Mr. Luke Bates for his careful reading of our manuscript. This work was supported by the Creative Research Initiative Project program (2015R1A3A2066497) funded by the National Research Foundation of the Korean government.

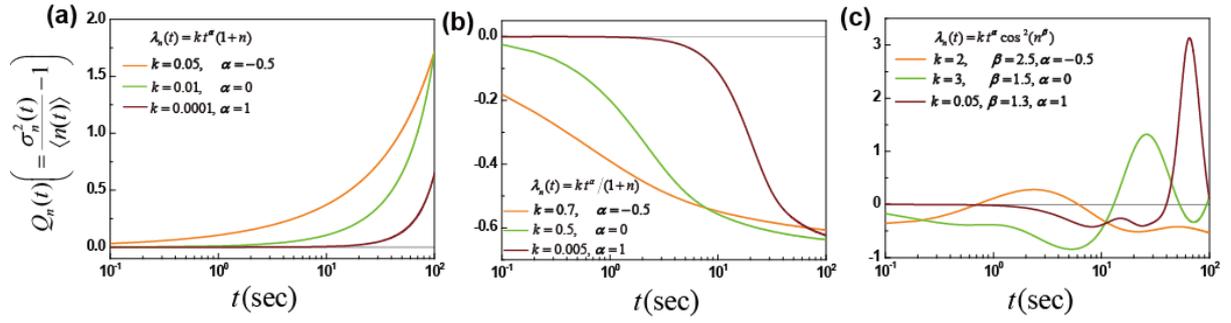

Fig1. Product number fluctuation with the reaction rates, $\lambda_n(t)$, which is a function of the number of product molecules. (a) The reaction rate, $\lambda_n(t) = k t^\alpha (1+n)$ $(k>0, \alpha>-1)$, is an increasing function of the product number. The Mandel's Q parameter of the product number always shows super-Poissonian characteristics $(Q_n(t)>0)$ over all time domains, regardless of $\alpha$. (b) $\lambda_n(t) = k t^\alpha /(1+n)$ $(k>0, \alpha>-1)$ decreases with the product number. Product number fluctuation always shows sub-Poissonian characteristics $(Q_n(t)<0)$ over all time domains. (c) $\lambda_n(t) = k t^\alpha \cos^2(n^\beta)$ $(k>0, \alpha>-1, \beta \text{ on } [-\infty,\infty])$ oscillates with the number of product molecules. The Mandel's Q parameter of the product number, $Q_n(t)$, oscillates with time.

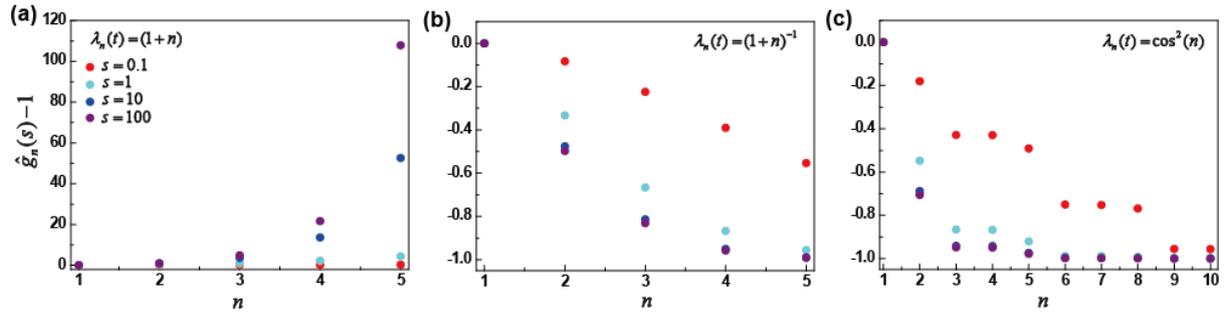

Fig.2. Nonrenewal quotient for three models of $\lambda_n(t)$. (a) $\lambda_n(t) = 1+n$. Nonrenewal quotient $\hat{g}_n(s) - 1$ sharply increases with the product number. (b) $\lambda_n(t) = (1+n)^{-1}$. The nonrenewal quotient, $\hat{g}_n(s) - 1$, decreases and approaches -1 as the number of product molecules increases. (c) $\lambda_n(t) = \cos^2(n)$. The slope of nonrenewal quotient shows oscillatory behavior unlike $\lambda_n(t) = 1+n$ and $\lambda_n(t) = (1+n)^{-1}$. The nonrenewal quotient for these three models deviates farther from zero as the Laplace variable, s, in $\hat{g}_n(s)$ becomes larger with the fixed number of product molecules, n.

# Supplemental Material

**Exploiting product molecule number to consider reaction rate fluctuation in elementary reactions**

Seong Jun Park[1]

[1] National CRI-Center for Chemical Dynamics in Living Cells, Chung-Ang University, Seoul 06974, Korea.

# Table of Contents

## SUPPLEMENTARY METHODS

# SUPPLEMENTARY METHODS

**Supplementary Method I |** Representation of $P_n(t)$ as a function of $\lambda_n(t)$

Eq. (2) is the first-order linear differential equation. If we assume that product number is zero at time zero, the initial condition of the probability of the product number being $n$ at time 0 is given by Kronecker delta $\delta_{n0}$ ($P_n(0) = \delta_{n0}$). Thus, under the initial condition,

$$P_n(t) = e^{-r_n(t)} \left( \int_0^t P_{n-1}(\tau) \lambda_{n-1}(\tau) e^{r_n(\tau)} d\tau + \delta_{n0} \right) \quad \text{(M1-1)}$$

where $r_n(t)$ is the integrating factor, $r_n(t) \equiv \int_0^t \lambda_n(\tau) d\tau$. Even though we obtain Eq. (M1-1), we have difficulty to calculate $P_n(t)$ for sufficiently large $n$ since the hierarchical structure of $P_n(t)$.

We will introduce a more useful formula than Eq. (M1-1) when we try to calculate the mean and variance of the number of product molecules. Because Eq. (M1-1) for $P_n(t)$ has a hierarchical structure, the largest trouble in Eq. (M1-1) is the time-consuming effort needed to be done in order to obtain the mean and variance of the number of product molecules. We re-derive $P_n(t)$ just as a function of the reaction rate when the number of product is $n$ at time $t$.

We begin with Eq. (M1-1). The so-called survival probability $P_0(t) = e^{-r_0(t)}$ where $r_n(t) = \int_0^t \lambda_n(\tau) d\tau$ for $n=0,1,2,\cdots$. And then

$$\begin{aligned} P_1(t) &= e^{-r_1(t)} \int_0^t \lambda_0(\tau) e^{r_1(\tau) - r_0(\tau)} d\tau \\ &= -e^{-r_1(t)} \int_0^t e^{r_1(\tau)} \frac{de^{-r_0(\tau)}}{d\tau} d\tau \end{aligned} \quad \text{(M1-2)}$$

Integrating by parts on the right side of Eq. (M1-2), we obtain

$$\begin{aligned} P_1(t) &= -e^{-r_1(t)} \left\{ \left[ e^{r_1(\tau) - r_0(\tau)} \right]_0^t - \int_0^t \lambda_1(\tau) e^{r_1(\tau) - r_0(\tau)} d\tau \right\} \\ &= e^{-r_1(t)} - e^{-r_0(t)} + e^{-r_1(t)} \int_0^t \lambda_1(\tau) e^{r_1(\tau) - r_0(\tau)} d\tau \\ &= (1 + q_1(t)) e^{-r_1(t)} - e^{-r_0(t)} \end{aligned} \quad \text{(M1-3)}$$

where $q_1(t) = \int_0^t \lambda_1(\tau) e^{r_1(\tau) - r_0(\tau)} d\tau$. $P_2(t)$ can be expressed by using Eq. (M1-3).

$$\begin{aligned} P_2(t) &= e^{-r_2(t)} \int_0^t e^{r_2(\tau)} \lambda_1(\tau) P_1(\tau) d\tau \\ &= e^{-r_2(t)} \int_0^t e^{r_2(\tau)} \lambda_1(\tau) \left( (1 + q_1(\tau)) e^{-r_1(\tau)} - e^{-r_0(\tau)} \right) d\tau \\ &= -e^{-r_2(t)} \int_0^t (1 + q_1(\tau)) e^{r_2(\tau)} \frac{de^{-r_1(\tau)}}{d\tau} + \lambda_1(\tau) e^{r_2(\tau) - r_0(\tau)} d\tau \\ &= -e^{-r_2(t)} \left[ (1 + q_1(\tau)) e^{r_2(\tau) - r_1(\tau)} \Big|_{\tau=0}^{\tau=t} \right. \\ &\quad \left. - \int_0^t e^{r_2(\tau)} \left\{ \left( \frac{dq_1(\tau)}{d\tau} + (1 + q_1(\tau)) \lambda_2(\tau) \right) e^{-r_1(\tau)} - \lambda_1(\tau) e^{-r_0(\tau)} \right\} d\tau \right] \end{aligned} \quad \text{(M1-4)}$$

Because $\dfrac{dq_1(\tau)}{d\tau} e^{-r_1(\tau)} = \lambda_1(\tau) e^{-r_0(\tau)}$, Eq. (M1-4) becomes

$$\begin{aligned} P_2(t) &= e^{-r_2(t)} \left( 1 - (1 + q_1(t)) e^{r_2(t) - r_1(t)} + \int_0^t (1 + q_1(\tau)) \lambda_2(\tau) e^{r_2(\tau) - r_1(\tau)} d\tau \right) \\ &= (1 + q_2(t)) e^{-r_2(t)} - (1 + q_1(t)) e^{-r_1(t)} \end{aligned} \quad \text{(M1-5)}$$

Where $q_2(t) = \int_0^t (1 + q_1(\tau)) \lambda_2(\tau) e^{r_2(\tau) - r_1(\tau)} d\tau$. In similar way, we obtain

$$\begin{aligned} P_3(t) &= e^{-r_3(t)} \int_0^t e^{r_3(\tau)} \lambda_2(\tau) P_2(\tau) d\tau \\ &= e^{-r_3(t)} \int_0^t e^{r_3(\tau)} \lambda_2(\tau) \left( (1 + q_2(\tau)) e^{-r_2(\tau)} - (1 + q_1(\tau)) e^{-r_1(\tau)} \right) d\tau \\ &= -e^{-r_3(t)} \int_0^t e^{r_3(\tau)} \left( (1 + q_2(\tau)) \frac{de^{-r_2(\tau)}}{d\tau} + (1 + q_1(\tau)) \lambda_2(\tau) e^{-r_1(\tau)} \right) d\tau \\ &= -e^{-r_3(t)} \left[ (1 + q_2(\tau)) e^{r_3(\tau) - r_2(\tau)} \Big|_{\tau=0}^{\tau=t} \right. \\ &\quad \left. - \int_0^t e^{r_3(\tau)} \left\{ \left( \frac{dq_2(\tau)}{d\tau} + (1 + q_2(\tau)) \lambda_3(\tau) \right) e^{-r_2(\tau)} - \lambda_2(\tau) (1 + q_1(\tau)) e^{-r_1(\tau)} \right\} d\tau \right] \end{aligned} \quad \text{(M1-6)}$$

Because $\dfrac{dq_2(\tau)}{d\tau} e^{-r_2(\tau)} = (1 + q_1(\tau)) \lambda_2(\tau) e^{-r_1(\tau)}$, Eq.(M1-6) becomes

$$\begin{aligned} P_3(t) &= e^{-r_3(t)} \left( 1 - (1 + q_2(t)) e^{r_3(t) - r_2(t)} + \int_0^t (1 + q_2(\tau)) \lambda_3(\tau) e^{r_3(\tau) - r_2(\tau)} d\tau \right) \\ &= (1 + q_3(t)) e^{-r_3(t)} - (1 + q_2(t)) e^{-r_2(t)} \end{aligned} \quad \text{(M1-7)}$$

Where $q_3(t) = \int_0^t (1+q_2(\tau))\lambda_3(\tau)e^{H_3(\tau)-H_2(\tau)}d\tau$. As continuing this process successively, one finds Eq. (M1-1) is transformed as follows.

$$P_n(t) = e^{-r_n(t)}(1+q_n(t)) - e^{-r_{n-1}(t)}(1+q_{n-1}(t))$$
$$= z_n(t) - z_{n-1}(t)$$
(M1-8)

for $n=0,1,2,\cdots$. Where $z_n(t) = e^{-r_n(t)}(1+q_n(t))$, $q_n(t) = \int_0^t (1+q_{n-1}(\tau))\lambda_n(\tau)e^{r_n(\tau)-r_{n-1}(\tau)}d\tau$, $r_n(t) = \int_0^t \lambda_n(\tau)d\tau$, and $q_0(t) = 0$. The physical interpretation of $z_n(t)$ is discussed in Supplementary Method IV.

**Supplementary Method II | Derivation of Eqs. (3) and (4)**

For the reaction rate $\lambda_n(t)$ when the number of product is $n$ at time $t$, the probability density $P_n(t)$ of having $n$ product molecules at time $t$ satisfies Eq. (2). The $m$-th moment of a physical quantity $x$ is defined as $\langle x^m(t) \rangle \equiv \sum_{n=0}^{\infty} x^m P_n(t)$, with $x \in \{n, \lambda_n(t), n\lambda_n(t)\}$.

Applying $\sum_{n=0}^{\infty} n$ to the both sides of Eq. (2), can result to the following equation:

$$\begin{aligned}
\frac{\partial \langle n(t) \rangle}{\partial t} &= \sum_{n=0}^{\infty} n \lambda_{n-1}(t) P_{n-1}(t) - \langle n \lambda_n(t) \rangle \\
&= \sum_{l=-1}^{\infty} (l+1) \lambda_l(t) P_l(t) - \langle n \lambda_n(t) \rangle \\
&= \sum_{n=0}^{\infty} (n+1) \lambda_n(t) P_n(t) - \langle n \lambda_n(t) \rangle \\
&= \sum_{n=0}^{\infty} \lambda_n(t) P_n(t) = \langle \lambda_n(t) \rangle
\end{aligned} \quad \text{(M2-1)}$$

because $\lambda_n(t)$ and $P_n(t)$ cannot be defined when $n$ is a negative integer. If we assume that the number of product molecules is zero at time zero, the mean number of the product molecules is the definite integral from zero to time $t$ of $\lambda_n(\tau)$ with respect to $\tau$, Eq. (3).

To find the variance of the number of product molecules, we first must find its second moment. Applying $\sum_{n=0}^{\infty} n^2$ to the both sides of Eq. (2), then we get:

$$\begin{aligned}
\frac{\partial \langle n^2(t) \rangle}{\partial t} &= \sum_{n=0}^{\infty} n^2 \lambda_{n-1}(t) P_{n-1}(t) - \langle n^2 \lambda_n(t) \rangle \tau \\
&= \sum_{l=-1}^{\infty} (l+1)^2 \lambda_l(t) P_l(t) - \langle n^2 \lambda_n(t) \rangle \\
&= \sum_{n=0}^{\infty} (n+1)^2 \lambda_n(t) P_n(t) - \langle n^2 \lambda_n(t) \rangle \\
&= \sum_{n=0}^{\infty} (2n+1) \lambda_n(t) P_n(t) \\
&= 2 \langle n \lambda_n(t) \rangle + \frac{\partial \langle n(t) \rangle}{\partial t}
\end{aligned} \quad \text{(M2-2)}$$

We use Eq. (M2-1) to obtain Eq. (M2-2). By integrating the both sides of Eq. (M2-2) with

respect to time $t$, one obtains

$$\langle n^2(t) \rangle = \langle n(t) \rangle + 2 \int_0^t \langle n \lambda_n(\tau) \rangle d\tau \tag{M2-3}$$

Using Eq. (M2-3), we find that the variance the number of product molecules is given by

$$\sigma_n^2(t) = \langle n^2(t) \rangle - \langle n(t) \rangle^2$$
$$= \langle n(t) \rangle (1 - \langle n(t) \rangle) + 2 \int_0^t \langle n \lambda_n(\tau) \rangle d\tau \tag{M2-4}$$

We can rewrite $\langle n(t) \rangle^2 = 2 \int_0^t \langle n(\tau) \rangle \frac{\partial \langle n(\tau) \rangle}{\partial \tau} d\tau$, substituting this into Eq. (M2-4),

$$\sigma_n^2(t) = \langle n(t) \rangle + 2 \int_0^t \langle n \lambda_n(\tau) \rangle - \int_0^t \langle n(\tau) \rangle \frac{\partial \langle n(\tau) \rangle}{\partial \tau} d\tau$$
$$= \langle n(t) \rangle + 2 \int_0^t \langle n \lambda_n(\tau) \rangle - \langle n(\tau) \rangle \langle \lambda_n(\tau) \rangle d\tau \tag{M2-5}$$

The integrand in the right-hand side of Eq. (M2-5) is the covariance of $n$ and $\lambda_n(t)$, $Cov(n, \lambda_n(t)) \equiv \langle n \lambda_n(\tau) \rangle - \langle n(\tau) \rangle \langle \lambda_n(\tau) \rangle$. According to the statistics, the covariance $Cov(n, \lambda_n(t))$ of $n$ and $\lambda_n(t)$ is the correlation coefficient $\rho(t)$ of the two variables $n$ and $\lambda_n(t)$ multiplied by the product of their standard deviations $\sigma_n(t) \sigma_{\lambda_n}(t)$. That is, $Cov(n, \lambda_n(t)) = \rho(t) \sigma_n(t) \sigma_{\lambda_n}(t)$. In addition, the correlation coefficient $\rho(t)$ ranges from -1 to 1. The positive value of $\rho(t)$ implies that the reaction rate $\lambda_n(t)$ when the number of product molecules is $n$ at time $t$ increases as the number of product molecules $n$ increases. The negative value of $\rho(t)$ implies that $\lambda_n(t)$ decreases as $n$ increases. The zero correlation implies that $\lambda_n(t)$ is independent of $n$, putting it differently, $\lambda_n(t)$ does not depend on $n$. Substituting $\langle n \lambda_n(\tau) \rangle - \langle n(\tau) \rangle \langle \lambda_n(\tau) \rangle = \rho(\tau) \sigma_n(\tau) \sigma_{\lambda_n}(\tau)$ into Eq. (M2-5), we finally obtain

$$\sigma_n^2(t) = \langle n(t) \rangle + 2 \int_0^t \rho(\tau) \sigma_n(\tau) \sigma_{\lambda_n}(\tau) d\tau \tag{M2-6}$$

One can easily derive Eq. (4) by arranging Eq. (M2-6).

**Supplementary Method III |** The product number fluctuation with the reaction rate being not dependent on the number of the product molecules

Eq. (4) demonstrates that the counting statistics of the number of product molecules is Poissonian character because the correlation coefficient $\rho(t)$ is equal to zero or that the reaction rate $\lambda_n(t)$ does not depend on the number of product molecules $n$. We can prove that fact in detail. Let the reaction rate $\lambda_n(t) = \lambda(t)$ and $r_n(t) = r(t) = \int_0^t \lambda(\tau) d\tau$ since the reaction rate does not depend on the number of product molecules. In other words, the reaction rate is just a function of time $t$ or a constant. Substituting $\lambda_n(t) = \lambda(t)$ and $r_n(t) = r(t) = \int_0^t \lambda(\tau) d\tau$ into Eq. (M1-8) gives us

$$P_0(t) = e^{-r(t)} \qquad (M3\text{-}1)$$

The same process is continued to get $P_1(t)$, $P_2(t)$, and $P_3(t)$.

$$\begin{aligned} P_1(t) &= e^{-r(t)}(1+q_1(t)) - e^{-r(t)} \\ &= q_1(t) e^{-r(t)} \\ &= r(t) e^{-r(t)} \end{aligned} \qquad (M3\text{-}2)$$

$$\begin{aligned} P_2(t) &= e^{-r(t)}\left(1+q_2(t) - (1+q_1(t))\right) \\ &= e^{-r(t)}\left(q_2(t) - r(t)\right) \\ &= e^{-r(t)}\left(\int_0^t (1+r(\tau))\lambda(\tau) d\tau - r(t)\right) \\ &= \frac{r(t)^2 e^{-r(t)}}{2} \end{aligned} \qquad (M3\text{-}3)$$

$$\begin{aligned} P_3(t) &= e^{-r(t)}\left(1+q_3(t) - (1+q_2(t))\right) \\ &= e^{-r(t)}\left(q_3(t) - \left(r(t) + \frac{r(t)^2}{2}\right)\right) \\ &= e^{-r(t)}\left(\int_0^t \left(1+r(t) + \frac{r(t)^2}{2}\right)\lambda(\tau) d\tau - \left(r(t) + \frac{r(t)^2}{2}\right)\right) \\ &= \frac{r(t)^3 e^{-r(t)}}{2 \cdot 3} \end{aligned} \qquad (M3\text{-}4)$$

We are going to form a conclusion:

$$P_n(t) = \frac{r(t)^n e^{-r(t)}}{n!} \qquad (\text{M3-5})$$

for $n=0,1,2,\cdots$. Eq. (M3-5) is Poisson distribution with mean and variance, $r(t)$. Thus, Mandel's Q parameter of product number, $\frac{\sigma_n^2(t)}{\langle n(t) \rangle} - 1$, is equal to zero because the variance of Eq. (M3-5) is the same as its mean, $\langle n(t) \rangle = \langle \sigma_n^2(t) \rangle = r(t)$.

**Supplementary Method IV | The relationship of $P_n(t)$ to $\varphi_n(t)$**

In this section, we will present the relationship between , $P_n(t)$ , the probability that the number of product molecules is *n* at time *t* , and $\varphi_n(t)$ , the probability density function of the time required to produce *n* product molecules. Using this relationship we will explain the physical meaning of $z_n(t) = e^{-r_n(t)}(1 + q_n(t))$, where $q_n(t) = \int_0^t (1 + q_{n-1}(\tau))\lambda_n(\tau) e^{r_n(\tau) - r_{n-1}(\tau)} d\tau$, $r_n(t) = \int_0^t \lambda_n(\tau) d\tau$, and $q_0(t) = 0$.

We will now let $T_n$ denote the waiting time until the number of product molecules reaches *n*, or the reaction occurs *n* times in an elementary reaction. The cumulative distribution function of $T_n$ when $t \geq 0$ is given by

$$F(t) = \Pr(T_n \leq t) = 1 - \Pr(T_n > t)$$
$$= 1 - \Pr(\text{the number of product molecules is less than } n \text{ in } [0,t]) \quad \text{(M4-1)}$$

Where $\Pr(x)$ is the probability of *x*. $\Pr(T_n \leq t)$ is the probability of $T_n \leq t$. In other words, $\Pr(T_n \leq t)$ is the probability that the number of product molecules is greater than or equal to *n* in time [0,*t*]. Let $\varphi_1(t)$ be the probability density function of time *t* required to produce one product molecule. Using the normalization condition, $\sum_{n=0}^{\infty} P_n(t) = 1$, we obtain

$$\Pr(T_1 \leq t) = \int_0^t \varphi_1(\tau) d\tau$$
$$= \sum_{n=1}^{\infty} P_n(t) = 1 - P_0(t) \quad \text{(M4-2)}$$

We can find $P_0(t) = 1 - \int_0^t \varphi_1(\tau) d\tau$, which is the so-called survival probability since $P_0(t)$ expresses the probability that reactant survives without a reaction in time [0,*t*]. Let $\varphi_2(t)$ be the probability density function of time *t* required to produce two product molecules. We then get

$$\Pr(T_2 \leq t) = \int_0^t \varphi_2(\tau) d\tau$$
$$= \sum_{n=2}^{\infty} P_n(t) = 1 - P_0(t) - P_1(t) \quad \text{(M4-3)}$$

Thus, $P_1(t) = \int_0^t \varphi_1(\tau) - \varphi_2(\tau) d\tau$. In the same way, Let $\varphi_3(t)$ be the probability density function of time $t$ required to produce three product molecules. Then,

$$\Pr(T_3 \leq t) = \int_0^t \varphi_3(\tau) d\tau \\ = \sum_{n=3}^{\infty} P_n(t) = 1 - \sum_{n=0}^{2} P_n(t) \quad \text{(M4-4)}$$

We find $P_2(t) = \int_0^t \varphi_2(\tau) - \varphi_3(\tau) d\tau$. Proceeding by induction one finds

$$P_n(t) = \begin{cases} 1 - \int_0^t \varphi_1(\tau) d\tau & (n = 0) \\ \int_0^t \varphi_n(\tau) - \varphi_{n+1}(\tau) d\tau & (n = 1, 2, 3, \cdots) \end{cases} \quad \text{(M4-5)}$$

Eq. (M4-5) has an interesting interpretation. It states that $P_n(t)$ equals subtracting the probability that the number of product molecules is greater than or equal to $n+1$ in time $[0,t]$ from the probability that the number of product molecules is greater than or equal to $n$ in time $[0,t]$.

The mean and second moment of the number of product molecules can be also obtained from Eq. (M4-5),

$$\langle n(t) \rangle = \sum_{n=0}^{\infty} n P_n(t) = \sum_{n=0}^{\infty} n \int_0^t \varphi_n(\tau) - \varphi_{n+1}(\tau) d\tau \\ = \sum_{n=1}^{\infty} n \int_0^t \varphi_n(\tau) - \varphi_{n+1}(\tau) d\tau \quad \text{(M4-6)} \\ = \sum_{n=1}^{\infty} \int_0^t \varphi_n(\tau) d\tau$$

Eq. (M4-6) tells us that the mean number of product molecules is the sum of all $\varphi_n(t)$, from $n$ equals one to infinity. From Eq. (M4-5) it follows that

$$\begin{aligned}
\langle n^2(t)\rangle &= \sum_{n=0}^{\infty} n^2 P_n(t) = \sum_{n=0}^{\infty} n^2 \int_0^t \varphi_n(\tau) - \varphi_{n+1}(\tau)\, d\tau \\
&= \sum_{n=1}^{\infty} n^2 \int_0^t \varphi_n(\tau) - \varphi_{n+1}(\tau)\, d\tau \\
&= \sum_{n=1}^{\infty} (2n-1) \int_0^t \varphi_n(\tau)\, d\tau \\
&= 2\sum_{n=1}^{\infty} n \int_0^t \varphi_n(\tau)\, d\tau - \langle n(t)\rangle
\end{aligned} \qquad \text{(M4-7)}$$

And so, since $\sigma_n^2(t) = \langle n^2(t)\rangle - \langle n(t)\rangle^2$, it follows that

$$\begin{aligned}
\sigma_n^2(t) &= 2\sum_{n=1}^{\infty} n \int_0^t \varphi_n(\tau)\, d\tau - \langle n(t)\rangle - \langle n(t)\rangle^2 \\
&= \langle n(t)\rangle + 2\int_0^t \left( \sum_{n=1}^{\infty} (n-1)\varphi_n(\tau) - \langle n(\tau)\rangle \frac{d\langle n(\tau)\rangle}{d\tau} \right) d\tau \\
&= \langle n(t)\rangle + 2\int_0^t \left( \sum_{n=1}^{\infty} (n-1)\varphi_n(\tau) - \langle n(\tau)\rangle \sum_{n=1}^{\infty} \varphi_n(\tau) \right) d\tau
\end{aligned} \qquad \text{(M4-8)}$$

using $\langle n(t)\rangle^2 = 2\int_0^t \left( \langle n(\tau)\rangle \dfrac{d\langle n(\tau)\rangle}{d\tau} \right) d\tau$.

To gain insight into the meaning of $z_n(t) = e^{-H_n(t)}(1 + q_n(t))$, we combine Eq. (M1-8) with Eq. (M4-5). Then, one can obtain

$$P_n(t) = z_n(t) - z_{n-1}(t) = \begin{cases} 1 - \int_0^t \varphi_1(\tau)\, d\tau & (n=0) \\ \int_0^t \varphi_n(\tau) - \varphi_{n+1}(\tau)\, d\tau & (n=1,2,3,\cdots) \end{cases} \qquad \text{(M4-9)}$$

Where $z_n(t) = e^{-r_n(t)}(1 + q_n(t))$, $r_n(t) = \int_0^t \lambda_n(\tau)\, d\tau$, $q_n(t) = \int_0^t (1 + q_{n-1}(\tau))\lambda_n(\tau) e^{r_n(\tau) - r_{n-1}(\tau)}\, d\tau$ and $q_0(t) = 0$. From Eq. (M4-9), this means that $P_0(t) = z_0(t) = 1 - \int_0^t \varphi_1(\tau)\, d\tau$. Substituting $n=1$ into Eq. (M4-9), we find

$$\begin{aligned}
P_1(t) &= z_1(t) - z_0(t) \\
&= z_1(t) - \left(1 - \int_0^t \varphi_1(\tau)\, d\tau\right) \\
&= \int_0^t \varphi_1(\tau) - \varphi_2(\tau)\, d\tau
\end{aligned}$$

Thus

$$z_1(t) = 1 - \int_0^t \varphi_2(\tau) d\tau \tag{M4-10}$$

Then we carry out the next similar computation. Substitute *n*=2 into Eq. (M4-9) and one easily finds

$$P_2(t) = z_2(t) - z_1(t)$$
$$= z_2(t) - \left(1 - \int_0^t \varphi_2(\tau) d\tau\right)$$
$$= \int_0^t \varphi_2(\tau) - \varphi_3(\tau) d\tau$$

Thus

$$z_2(t) = 1 - \int_0^t \varphi_3(\tau) d\tau \tag{M4-11}$$

At about this point one can suspect a pattern emerging. It appears that

$$z_n(t) = 1 - \int_0^t \varphi_{n+1}(\tau) d\tau = e^{-r_n(t)}(1 + q_n(t)) \tag{M4-12}$$

for $n = 0, 1, 2, \cdots$, Where $q_n(t) = \int_0^t (1 + q_{n-1}(\tau))\lambda_n(\tau) e^{r_n(\tau) - r_{n-1}(\tau)} d\tau$, $r_n(t) = \int_0^t \lambda_n(\tau) d\tau$, and $q_0(t) = 0$. Since the integral $\int_0^t \varphi_{n+1}(\tau) d\tau$ means the probability that the number of product molecules is greater than or equal to *n*+1 in time [0,*t*], $z_n(t)$ is therefore interpreted as the probability that the number of product molecules is less than or equal to *n* in time [0,*t*].

One may find that $\lambda_n(t) P_n(t) = \varphi_{n+1}(t)$ by comparing Eq. (3) and Eq. (M2-5) with Eq. (M4-6) and Eq. (M4-8) for $n = 0, 1, 2, \cdots$. We shall prove it closely. We know from Eq. (M4-12) that the derivative of $z_n(t)$ with respect to *t* equals $\varphi_{n+1}(t)$

$$-\frac{\partial z_n(t)}{\partial t} = \varphi_{n+1}(t) \tag{M4-13}$$

for *n*=0,1,2,⋯. Using $z_n(t) = e^{-r_n(t)}(1 + q_n(t))$, we can find then see

$$\begin{aligned}
\varphi_{n+1}(t) &= -\frac{\partial}{\partial t}\left(e^{-r_n(t)}\left(1+q_n(t)\right)\right) \\
&= -\left\{-\lambda_n(t)e^{-r_n(t)}\left(1+q_n(t)\right)+\frac{\partial q_n(t)}{\partial t}e^{-r_n(t)}\right\} \\
&= -\left\{-\lambda_n(t)e^{-r_n(t)}\left(1+q_n(t)\right)+\lambda_n(t)e^{-r_{n-1}(t)}\left(1+q_{n-1}(t)\right)\right\} \\
&= \lambda_n(t)e^{-r_n(t)}\left(1+q_n(t)\right)-\lambda_n(t)e^{-r_{n-1}(t)}\left(1+q_{n-1}(t)\right) \\
&= \lambda_n(t)\left(z_n(t)-z_{n-1}(t)\right) \\
&= \lambda_n(t)P_n(t)
\end{aligned} \qquad \text{(M4-14)}$$

If the reaction rate $\lambda_n(t)$ with the number of product molecules $n$ at time $t$ is known, we can predict the product number fluctuation and the probability density function of the time required to produce $n$ product molecules.

Using Eq. (M4-9), we are then able to rewrite the mean and the variance of the product number, originally given as Eqs. (3) and (M2-5), respectively as

$$\langle n(t) \rangle = \sum_{n=1}^{\infty}\left(1-z_{n-1}(t)\right) \qquad \text{(M4-15)}$$

$$\sigma_n^2(t) = 2\sum_{n=1}^{\infty}\left(n\left(1-z_{n-1}(t)\right)\right)-\langle n(t)\rangle\left(1+\langle n(t)\rangle\right). \qquad \text{(M4-16)}$$

Eqs. (M4-15) and (M4-16) are more comfortable than Eqs. (3) and (M2-5) in terms of calculation of product number counting statistics. Due to the fact that $z_n(t)$ tends to go to zero as $n$ goes to infinity, Eqs. (M4-15) and (M4-16) are in fact more convenient in calculating the mean and variance of the product number.

**Supplementary Method V | Comparison of Eq. (4) and the Chemical Fluctuation Theorem**

Park *et al*. present the Chemical Fluctuation Theorem (CFT), a general relationship between the fluctuation in the number of product molecules and the dynamics of product creation and annihilation processes [1]. From comparison of Eq. (4) and the CFT without product annihilation process, one can easily find this relation

$$Cov(n, \lambda_n(t)) = \int_0^t \langle \delta R(t) \delta R(x) \rangle dx \tag{M5-1}$$

the covariance of $n$, product number, and $\lambda_n(t)$, the reaction rate when the product number is $n$ at time $t$, $Cov(n, \lambda_n(t)) \equiv \langle n\lambda_n(\tau) \rangle - \langle n(\tau) \rangle \langle \lambda_n(\tau) \rangle$,

$\langle \delta R(t_2) \delta R(t_1) \rangle = \langle R(t_2) R(t_1) \rangle - \langle R(t_2) \rangle \langle R(t_1) \rangle$ is the time correlation function (TCF) of the reaction rate.

When the reaction process is a stationary process, Eq. (M5-1) becomes

$$\frac{dCov(n, \lambda_n(t))}{dt} = \langle \delta R(t) \delta R(0) \rangle \tag{M5-2}$$

In terms of $\varphi_n(t)$, the probability density function of the time required to produce $n$ product molecules, we can rewrite Eq. (M5-2) using $\langle n(t) \rangle = \langle R \rangle t$,

$$\langle \delta R(t) \delta R(0) \rangle = \sum_{n=0}^{\infty} n \frac{d\varphi_{n+1}(t)}{dt} - \langle R \rangle^2 \tag{M5-3}$$

where $\langle R \rangle = \lim_{t \to \infty} \sum_{n=1}^{\infty} \varphi_n(t)$. If the reaction process is a stationary renewal process, the Laplace transform of $\varphi_n(t)$ equals $\hat{\varphi}(s)^{n-1} \langle R \rangle \frac{1 - \hat{\varphi}(s)}{s}$, where $\hat{\varphi}(s)$ is the Laplace transform of $\varphi(t)$, the probability density function of reaction time interval. Thus, we obtain

$$\langle \delta \hat{R}(s) \delta R(0) \rangle = \langle R \rangle \left( \frac{\hat{\varphi}(s)}{1 - \hat{\varphi}(s)} - \frac{\langle R \rangle}{s} \right) \tag{M5-4}$$

where $\hat{f}(s)$ denotes the Laplace transform of $f(t)$. For non-stationary reaction process, we can find the TCF of the reaction rate by solving Eq. (M5-1) when $\varphi_n(t)$ is known.